\documentstyle[preprint,prl,aps,array,graphicx]{revtex}
\newcommand{\be}{\begin{equation}}
\newcommand{\ee}{\end{equation}}
\begin{document}
\title{$q$-Thermostatistics and the black-body radiation
problem}
\author{S. Mart\'{\i}nez
\thanks{
E-mail: martinez@venus.fisica.unlp.edu.ar}, F. Pennini\thanks{
E-mail: pennini@venus.fisica.unlp.edu.ar}, A. Plastino\thanks{
E-mail: plastino@venus.fisica.unlp.edu.ar}, and C. J.
Tessone\thanks{ E-mail: tessonec@venus.fisica.unlp.edu.ar}.}
\address{Instituto de F\'{\i}sica de La Plata, \\
 National University
La Plata (UNLP) and National Research Council (CONICET),\\C.C.
727, 1900 La Plata, Argentina. }\maketitle
\begin{abstract}
We give an exact information-theory treatment of the
$n$-dimensional black-body radiation process in a non-extensive
scenario. We develop a $q$ generalization of the laws of i) Stefan
Boltzmann, ii) Planck, and iii) Wien, and show that conventional,
canonical results are obtained at temperatures above $1 K$.
Classical relationships between radiation, pressure, and internal
energy are recovered (independently of the $q$-value). Analyzing
the particles' density for $q\approx 1$ we see that the
non-extensive parameter $q$ introduces a fictitious chemical
potential. We apply our results
 to experimental data on the cosmic microwave
background and reproduce it with acceptable accuracy for
different temperatures (each one associated to a particular $q$
value).
 \vspace{0.2 cm}


 \vspace{1
cm}
\end{abstract}


\newpage

\section{Introduction}

Black-body radiation studies constitute a milestone in the history
of Physics. Planck's law is satisfactorily accounted for by
recourse to Bose-Einstein statistics and has been experimentally
re-confirmed over and over for almost a century. However, in the
last decade  small deviations from this law have been detected in
the cosmic microwave radiation \cite{FIRAS}. In \cite{cn1}, its
authors advance the hypothesis that these deviations could have
arisen at the time of the matter-radiation de-coupling, and
associate them to a non-extensive statistics environment. They
argue in favor of such an hypothesis on the basis of the close
relation that exists between long-range interactions and
non-extensive scenarios \cite {pla2,pla4}. The concomitant
non-extensive  thermostatistical treatment \cite
{pla2,pla4,t01,t1,review,t2,t03,t3,pla1,pla3} is by now recognized
as a new paradigm for statistical mechanical considerations. It
revolves around the concept of Tsallis' information measure
$S_{q}$ \cite{t01}, a generalization of the logarithmic Shannon's
one. $S_q$ is parameterized by a real index $q$ and becomes
identical to Shannon's measure when $q=1$.

The study reported in \cite{cn1} employs the so-called
Curado-Tsallis un-normalized expectation values \cite{t3} in the
limit $q\rightarrow 1$. Nowadays, it is believed that the
Curado-Tsallis (CT) framework has been superseded by the so-called
normalized or Tsallis-Mendes-Plastino (TMP) one
\cite{TMP,pennini}, which seems to exhibit important advantages
\cite{review}. This normalized treatment, in turn, has been
considerably improved by the so-called ``Optimal Lagrange
Multipliers'' (OLM) approach \cite{OLM}. It is, then, natural to
revisit the problem from such a new viewpoint.

The OLM treatment can be recommended in view of the findings of
\cite{vander,Rama,temperature} regarding the particular nature of
the Lagrange multiplier associated to the temperature. Within an
OLM context, non-extensivity is restricted just to the entropy.
The internal energy remains extensive and
 the Lagrange multipliers conserve their traditional intensive character (allowing one to
identify them
 with their thermodynamic counterpart). Moreover, the OLM approach
  unifies Tsallis and R\'enyi
variational formalisms under
 a $q$-thermostatistics umbrella. The solutions to the
 concomitant OLM-Tsallis variational problem are also solutions
 for its  R\'enyi-counterpart \cite{Renyi}.
 Since R\'enyi's entropy is extensive, one easily understands
 thereby the OLM formalism's success in  reproducing classical thermodynamic
results in a simple and clean fashion \cite{virial,idealgas}.

We shall study here, from an OLM viewpoint, black-body radiation
in equilibrium within an enclosure of volume $V$, with the goal of
ascertaining the possible $q$-dependence of i) the Planck
spectrum, ii) Stefan-Boltzmann's law, and iii) Wien's one. We will
compare the ensuing results with those found in the literature
\cite{cn1,Lenzi,FA,Diego} (all of them under Curado-Tsallis
treatment) and with experimental data \cite{FIRAS}. Due to the
fact that most previous non-extensive treatments of quantal gases
employ the so-called Factorization Approach (FA), and that this
approximation underlies some of the preceding treatments of the
black-body radiation problem, we will translate the FA into the
language of a {\it normalized}-OLM Factorization Approach
(OLM-FA). This in turn provides one with a $q$-normalized
generalization of the particle's density expression for quantum
gases.

The paper is organized as follows: In Sect. \ref{solm} we present
a brief OLM primer. In Sect. \ref{cn} we obtain the OLM partition
function. Sect. \ref{SBL} is devoted to the Stefan-Boltzmann law.
We obtain the exact expression for the internal energy
(\ref{comp}) and find that the Stefan-Bolzmann law holds for all
temperatures except a small interval ($T\in[10^{-2}K,1K]$). We
also revisit pertinent antecedents in the literature of the
non-extensive approach to the black-body problem (\ref{history}).
In Sect. \ref{PWLs} we perform an exact calculation of the energy
density. Planck's law (\ref{PL}) seems to be valid everywhere,
with just slight variations in the shape of the associated curves.
Following \cite{cn1}, the formalism is applied to the cosmic
microwave radiation (\ref{cobe}) in order to search for deviations
from Planck's law. Using the maximua of the energy density curves,
we search for the $q$-generalization of Wien's law (\ref{WiL}) and
find good agreement, save for a small temperature interval
($T\in[10^{-4}K,10^{-2}K]$). Some conclusions are drawn in Sect.
\ref {conclusiones}. In Appendix \ref{OLMFA} we review the FA,
introduce our new OLM-FA technique and apply it to the black-body
radiation problem.

\section{Main results of the OLM formalism}
\label{solm}
 For a most general quantal treatment, in a
basis-independent way, consideration is required of the density
operator $\hat{\rho}$ that maximizes Tsallis' entropy
\begin{equation}
\frac{S_{q}}{k}=\frac{1-Tr(\hat{\rho}^{q})}{q-1}, \label{entropia}
\end{equation}
subject to the $M$ generalized expectation values $\left\langle
\widehat{O}%
_{j}\right\rangle_q $, where $\widehat{O}_{j}$ $(j=1,\ldots ,M)$
denote the relevant observables.

Tsallis' normalized probability distribution \cite{TMP} is
obtained by following the well known MaxEnt route \cite{katz}.
Instead of performing the variational treatment of
Tsallis-Mendes-Plastino (TMP) \cite{TMP}, we will pursue the
alternative path developed in \cite{OLM}. One maximizes Tsallis'
generalized entropy given by Eq. (\ref{entropia}) \cite{t01,t1,t2}
subject to the constraints \cite{t01,OLM}

\begin{eqnarray}
Tr(\hat{\rho}) &=&1 \\
Tr\left[ \hat{\rho}^q\left( \widehat{O}_j- \left\langle\widehat{O}%
_j\right\rangle _q\right) \right] &=&0,  \label{vinculos}
\end{eqnarray}
whose generalized expectation values \cite{TMP}

\begin{equation}  \label{gener}
\left\langle\widehat{O}_j \right\rangle _q = \frac{Tr(\hat \rho^q
\widehat{O}%
_j)}{Tr(\hat \rho^q)},
\end{equation}
are (assumedly) a priori known. In contrast with the TMP-instance
\cite{TMP}, the resulting density operator

\begin{equation}
\hat{\rho}=\frac{\hat{f}_q^{1/(1-q)}}{\bar{Z}_q},  \label{rho}
\end{equation}
is not self referential \cite{OLM}. In Eq. (\ref{rho}),
$\{\lambda_j\}$ ($j=1,...,M$) stands for the so-called ``optimal
Lagrange multipliers' set". The quantity $\hat f_q$ is known as
the configurational characteristic and takes the form \be
\hat{f}_q= 1-(1-q)\sum_j^M\,\lambda _j\left( \widehat{O}%
_j- \left\langle \widehat{O}_j\right\rangle _q\right),
\label{charact} \ee if its argument is positive, while otherwise
$\hat{f}_q=0$ (cut-off condition \cite{TMP}). Of course,
$\bar{Z}_q$ is the partition function

\begin{equation}
\bar{Z}_{q}=Tr \hat{f}_q^{1/(1-q)}.  \label{Zqp}
\end{equation}

It is shown in Ref. \cite{OLM} that, as a consequence of the
normalization condition, one has
\begin{equation}
{\mathcal R}_q\equiv Tr\hat{f}_q^{q/(1-q)}=\bar{Z}_q, \label{relac1}
\end{equation}
which allows one to write Tsallis's entropy as

\begin{equation}
S_{q}=k\;{\rm \ln }_{q}\bar{Z}_{q},  \label{S2}
\end{equation}
with ${\rm \ln }_{q}x=(1-x^{1-q})/(q-1)$. These results coincide
with those of TMP \cite{TMP}. Using Eq. (\ref{relac1}), the
connection between the OLM's set and the TMP's Lagrange
multipliers set can be written as \cite{OLM}

\be
\label{ljTMP} \lambda_j^{TMP}=\bar{Z}_q^{1-q} \lambda_j. \ee

 The TMP's Lagrange multipliers are not intensive quantities
 \cite{vander,temperature}. They do not, as a consequence, have a simple
 physical interpretation. To the contrary, the OLM multipliers
 {\em are} intensive \cite{vander,temperature}, a fact that can be easily
explained noticing that they
 are the natural Lagrange multipliers of a R\'enyi's variational approach \cite{Renyi}.

If we define now
\begin{equation}
\ln Z_{q}={\rm \ln }\bar{Z}_{q}-\sum_{j}\lambda _{j}\ \left\langle
\widehat{O%
}_{j}\right\rangle _{q},  \label{lnqz'}
\end{equation}
we are straightforwardly led to \cite{vander}
\begin{eqnarray}
\frac{\partial }{\partial \left\langle \widehat{O}%
_{j}\right\rangle _{q}}\left(\ln \bar{Z}_{q}\right)&=&\lambda
_{j},  \label{termo1} \\ \frac{\partial }{\partial \lambda
_{j}}\left( \ln Z_{q}\right) &=&- \left\langle
\widehat{O}_{j}\right\rangle _{q}.  \label{termo2}
\end{eqnarray}

Equations (\ref{termo1}) and (\ref{termo2}) are  fundamental
information theory (IT) relations for formulating Jaynes' version
of  statistical mechanics \cite{katz}.   Due to Eqs. (\ref{S2})
and (\ref{ljTMP}), the IT relation  (\ref{termo1}) leads
straightforwardly to the well known expression \cite{TMP}

\be
\frac{\partial }{\partial \left\langle \widehat{O}%
_{j}\right\rangle _{q}}\left(\frac{S_{q}}{k}\right)=\lambda^{TMP}
_{j}. \ee

\section{Partition function and Radiation pressure}
\label{cn}

We will introduce now the exact (OLM) black-body radiation
treatment. The chemical potential is taken, of course, equal to
zero (Grand Canonical ensemble with $\mu=0$) and, in looking for
the equilibrium properties we consider that the appropriate
thermodynamical variables are, as customary, the volume $V$,  and
the temperature $T$ \cite{Huang}. Consider first the standard
situation $q=1$.

The Hamiltonian of the electromagnetic field, in which there are $%
n_{{\bf k},{\bf \epsilon }}$ photons of momentum ${\bf k}$ and
polarization ${\bf %
\epsilon }$, is given by
\begin{equation}
\widehat{{\cal H}}=\sum_{_{k,\epsilon }}\hbar \omega
\hat{n}_{k,\epsilon }, \label{H}
\end{equation}
where the frequency is $\omega =c\left| {\bf k}\right| $
  and $n_{k,\epsilon }=0,1,2,\ldots $, with no
restrictions on $\left\{ n_{{\bf k},{\bf \epsilon} }\right\}.$

For a macroscopic volume $V$ the density of states $g$ in an $n$%
-dimensional space is $g_{n}(\omega )=A_{n}\omega ^{n-1}$, with

\begin{equation}
A_{n}=\frac{2\tau _{n}V}{(4\pi c^{2})^{n/2}\Gamma (n/2)},
\label{An}
\end{equation}
where $\tau _{n}=n-1$ is the number of linearly-independent polarizations.

The partition function
\begin{equation}
Z_{1}=Tr\left( e^{-\beta \widehat{{\cal H}}}\right)  \label{Z1}
\end{equation}
can be written as
\begin{equation}
Z_{1}=\exp \left\{ \int_{0}^{\infty }d\omega g_{n}(\omega )\ln
\left[ 1-\exp \left( -\beta \hbar \omega \right) \right] \right\}
=e^{\xi _{n}}, \label{Z1f}
\end{equation}
where
\begin{equation}
\xi _{n}=\frac{I_{n}A_{n}}{(\hbar \beta )^{n}},  \label{xin}
\end{equation}
and
\begin{equation}
I_{n}=-\int_{0}^{\infty }dxx^{n-1}\ln \left( 1-e^{-x}\right)
=\Gamma (n)\zeta (n+1),
\end{equation}
with $\zeta $ standing for the Riemann function and   $\Gamma $
for the Gamma function.

The OLM-Tsallis generalized configurational characteristic will be
(Cf. Eq. (\ref{charact}))

\be
\hat{f}_q=1-(1-q)\beta(\widehat{H}-U_{q}),\label{fq} \ee where
$\widehat{H}$ is the Hamiltonian given by Eq. (\ref{H}), $U_{q}$
is the mean energy introduced in Eq. (\ref{gener}), and
$\beta=1/kT $.

With the aim of calculating $\bar{Z}_{q}$ as defined by Eq.
(\ref{Zqp}), we follow the steps of Ref. \cite{Lenzi} and use the
integral (Gamma) definition given by the relation
\cite{Gradshteyn}

\begin{equation}  \label{pi}
b^{z-1}=\left\{
\begin{array}{cr}
 \frac{\Gamma (z)}{2\pi }\int\limits_{-\infty }^{\infty
}dt\frac{e^{(1+it)b}}{%
(1+it)^{z}}& \text{for }b>0 \\ 0 & \text{ for }b\leq 0,
\end{array}
\right.
\end{equation}
with $\mathop{\rm Re}(z)>0$, and $-\pi /2<\arg (a+it)<\pi /2.$

If we set $b={f}_q$ (the cut-off condition is naturally fulfilled
\cite{review}) and $z=1/(1-q)+1$ ($\mathop{\rm Re}(z)>0$, so that either $q>2$
or $q<1$), the generalized partition function adopts the appearance

\begin{equation}
\bar{Z}_{q}(U_q )=\int\limits_{-\infty }^{\infty }dtK_{q}(t)
Z_{1}(\tilde{\beta}),  \label{Zq}
\end{equation}
with $Z_{1}$ given by Eq. (\ref{Z1f}),
\begin{equation}
K_{q}(t)=\frac{\Gamma \left[ (2-q)/(1-q)\right] \exp
(1+it)e^{\tilde{\beta}U_{q}}}{2\pi (1+it)^{(2-q)/(1-q)}},
\label{Kq}
\end{equation}
and
\begin{equation}
\tilde{\beta}=(1+it)(1-q)\beta .  \label{bm}
\end{equation}

In order to evaluate the integral in Eq. (\ref{Zq}), we expand the
exponential  and obtain
\begin{equation}
\bar{Z}_{q}(U_q )=\frac{\Gamma [(2-q)/(1-q)]}{2\pi }
\sum\limits_{m=0}^{\infty }\frac{\xi _{n}^{m}}{m!}(1-q)^{-nm}
\int\limits_{-\infty }^{\infty }dt\frac{%
e^{(1+(1-q)\beta U_{q})(1+it)}}{(1+it)^{\frac{2-q}{1-q}+nm}},
\end{equation}
where $\xi _{n}$ is given by Eq. (\ref{xin}).

Using again Eq. (\ref{pi}), with $b=1+(1-q)\beta U_{q}$ and $%
z=(2-q)/(1-q)+nm$, we arrive at
\begin{equation}
\bar{Z}_q(U_q)=\Gamma \left[(2-q)/(1-q)\right]\left[ 1+(1-q)\beta
U_{q}\right] ^{1/(1-q)} \sum\limits_{m=0}^{\infty }B_m\Gamma^{-1}
\left[(2-q)/(1-q)+nm\right], \label{Zqf}
\end{equation}
where
\be
B_m=\frac{\xi _{n}^{m}}{m!}\frac{\left[ 1+(1-q)\beta U_{q}\right]
^{nm}}{%
(1-q)^{mn}}. \label{Bm} \ee Notice that an additional, cut-off-like condition
must be
considered, namely,  $$%
1+(1-q)\beta U_{q}>0,\,\,\, otherwise,\,\, \bar{Z}_{q}(U_q
)=0.$$

A similar path can be followed in order to obtain the quantity
${\cal R}_{q}$ (introduced in Eq. (\ref{relac1})), which will read

\begin{equation}
{\mathcal R}_q(U_q)=\Gamma \left[1/(1-q)\right]\left[ 1+(1-q)\beta
U_{q}\right] ^{q/(1-q)} \sum\limits_{m=0}^{\infty }B_m\Gamma^{-1}
\left[1/(1-q)+nm\right], \label{Cq}
\end{equation}
although in this case the allowed interval of $q$-values is
reduced to $0<q<1$. Note that this permissible interval of
$q$-values respects the new cut-off condition introduced above.

Once we have $\bar{Z}_q$, the radiation pressure is easily
obtained by applying Eq. (\ref{termo1}) to $(\beta P,V)$ \cite{Huang}, i.e.,

\be
\beta P=\frac{\partial}{\partial V}\ln
\bar{Z}_q=\frac{1}{V}\frac{\sum_m m B_m
\Gamma^{-1}\left((2-q)/(1-q)+nm\right)}{ \sum_m B_m
\Gamma^{-1}\left((2-q)/(1-q)+nm\right)}. \label{pressure} \ee

Due to the fact that we have set $\mu=0$, the quantity $\bar{Z}_q$
of Eq. (\ref{Zqf}) does not depend on the mean number of particles
$N_q$. One has

\be
\mu=\beta^{-1}\frac{\partial}{\partial N_q}\ln \bar{Z}_q=0.
\label{mu} \ee

\section{Stefan-Bolzmann's law}
\label{SBL}
\subsection{Exact OLM treatment}
\label{comp}

As in the previous section, the generalized internal energy of the
black-body radiation is obtained from Eq. (\ref{gener}) by
specializing the problem to the Grand Canonical ensemble with
$\mu=0$
\be
U_q={\mathcal R}_q^{-1} Tr \left( \hat{f}_q^{q/(1-q)}
\widehat H\right),
\ee
with $\hat{f}_q$ given by Eq. (\ref{fq}).

Consider now the trace's content. Using  (\ref{pi}) we can find

\be U_q={\mathcal R}_q^{-1} \frac{\Gamma [1/(1-q)]}{2\pi
}\int\limits_{-\infty }^{\infty } dt\frac{e^{(1+it)[1+(1-q)\beta
U_{q}]}}{(1+it)^{\frac{1}{1-q}}}Tr\left( Z_1(\tilde \beta)
\widehat{H}\right), \ee where $\tilde \beta$ is given by Eq.
(\ref{bm}). The permissible $q$-interval is $0<q<1$, due to
restrictions posed by the  Gamma integral representation.  Taking
advantage of the fact that

\begin{equation}
Tr\left[ Z_{1}(\tilde{\beta})\widehat{H}\right] =-\frac{\partial
Z_{1}(%
\tilde{\beta})}{\partial
\tilde{\beta}}=\frac{n}{\tilde\beta}\sum_m\frac{\xi_n^{m+1}(\tilde\beta)}{m!},
\label{traza}
\end{equation}
we obtain

\begin{equation}
U_q={\mathcal R}_q^{-1}\frac{n}{2\pi}\Gamma
\left(\frac1{1-q}\right) \frac{\xi_n}{(1-q)^{n+1}\beta}
\sum\limits_{m=0}^{\infty }%
\frac{\xi _{n}^{m}}{m!}\frac1{(1-q)^{nm}}\int\limits_{-\infty
}^{\infty
}dt\frac{%
e^{(1+it)[1+(1-q)\beta U_{q}]}}{(1+it)^{\frac{1}{1-q}+n(1+m)+1}},
\end{equation}
and, using again Eq. (\ref{pi}), we obtain the mean energy
expression
\begin{equation}
U_{q}=\frac{n\xi _{n}}{(1-q)^{n+1}\beta }\left[ 1+(1-q)\beta
U_{q}\right] ^{n+1}\frac{\sum\limits_{m=0}^{\infty} B_m
\Gamma^{-1} [1/(1-q)+n(m+1)+1]}{\sum\limits_{m=0}^{\infty }B_m
\Gamma^{-1} [1/(1-q)+nm]}. \label{Uqf}
\end{equation}

Notice that the Tsallis'cut-off condition $1+(1-q)\beta U_{q}>0$
is always satisfied. The series in Eq. (\ref{Uqf}) rapidly
converges on account of the exponential-like factors $B_m$ (Cf.
Eq. (\ref{Bm})), weighted by inverse Gamma functions.

Lenzi {\em et al.} solved the $q$-black-body radiation problem
{\em in exact} fashion in its  Curado-Tsallis un-normalized
version \cite{Lenzi}. Since their resulting internal energy is not
self referential (as it is in the TMP normalized instance) they
were able to describe the asymptotic behavior for $\beta$ in quite
simple terms. {\it Here} we need to perform a more detailed
analysis by considering  different possibilities for the form that
the product $\beta U_q$ may take as $\beta \rightarrow \infty$.
For instance,  if we assume that i) $\beta U_q$ goes over to a
constant or ii) it is not bounded, the limiting process leads to
incoherencies, while if we assume that, in Eq. (\ref{Uqf}), $\beta
U_q\rightarrow 0$ when $\beta \rightarrow \infty$, $U_q\propto
T^4$, as one has the right to expect.

As a consequence of the normalization condition given by Eq. (\ref{vinculos}),
we know that ${\mathcal R}_q=\bar{Z}_q$ (see Eq. (\ref{relac1})). This relation
allows us to look for alternative expressions for the relevant mean value. If
we evaluate $U_q$ in terms of $\bar{Z}_q$ we find

\be
U_q=\frac{n}{\beta}\frac{\sum\limits_{m=0}^{\infty} m B_m
\Gamma^{-1}\left((2-q)/(1-q)+nm\right)}{
\sum\limits_{m=0}^{\infty} B_m
\Gamma^{-1}\left((2-q)/(1-q)+nm\right)}. \label{Uq2} \ee

By inspection of Eq. (\ref{pressure}) one realizes  that the traditional
relation between internal energy and pressure still holds here

\be
P=\frac{1}{n}\frac{U_q}{V}. \ee

\begin{figure}[tbp]
\begin{center}
\includegraphics[width=11cm]{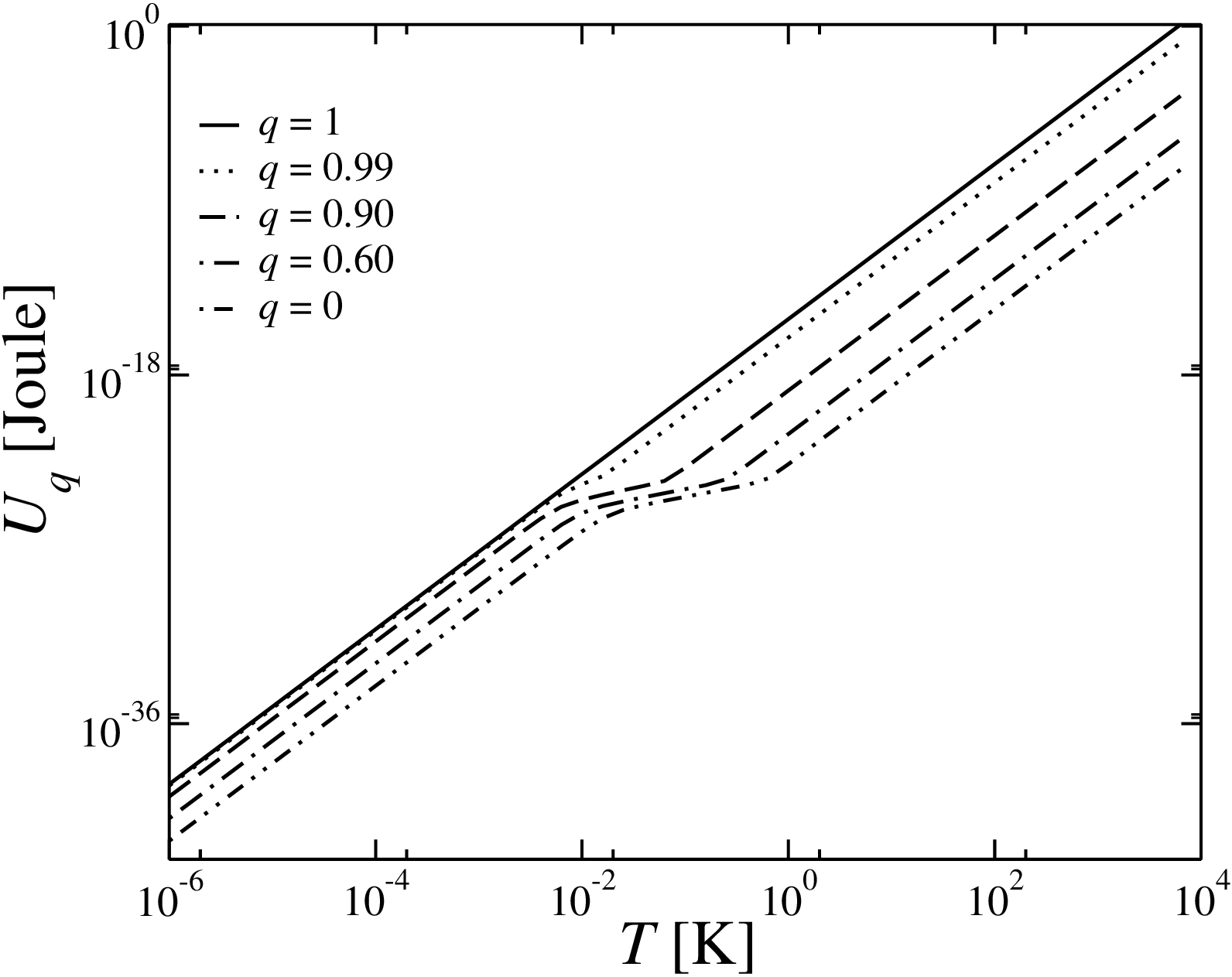}
\end{center}
\caption{Internal energy, $U_q$, of a three-dimensional system as
a function of Temperature $T=1/\beta$ for different values of the
non-extensivity parameter $q$ for a box whose volume is $1
\hbox{m}^3$. } \label{Uq-T}
\end{figure}

Eqs. (\ref{Uqf}) or (\ref{Uq2}) are recursive expressions that
give the OLM version of Stefan-Boltzmann's law, which, for the
$q=1$ case, reads

\be U=n\xi_n/\beta\propto T^{n+1}. \ee The present  equations are
to be tackled numerically. Fig. \ref{Uq-T} depicts $U_q$ as a
function of $T$ for different values of $q$ and $n=3$ in a
log-log scale, where $U_q$ has been evaluated from Eq.
(\ref{Uqf}). It is seen that Stefan-Boltzmann's law is reproduced
by our formalism for a wide range of $T$-values. Violations are
detected just for some special $T$-ranges (that depend on $q$),
within the interval $10^{-2}K<T<1K$.

\begin{figure}[tbp]
\begin{center}
\includegraphics[width=11cm]{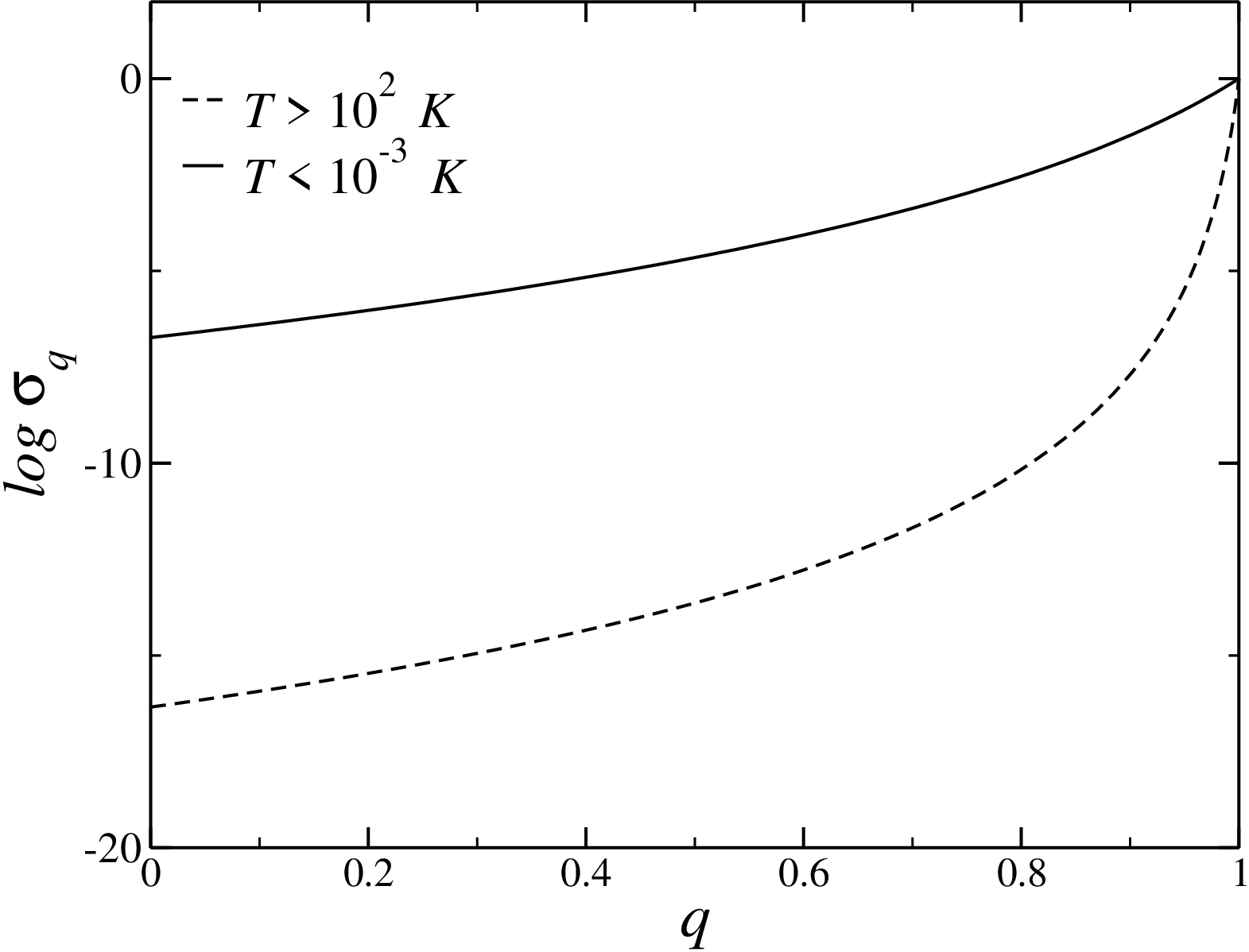}
\end{center}
\caption{Generalized Stefan constant $\sigma_q$ as a function the
index $q$. It is shown $\ln(\sigma_q/\sigma)\;vs\;q$ with
$\sigma=5.67051\cdot 10^{-8} \hbox{Watt}/ ( K^4  \hbox{Meter}^2)$
the Stefan constant for $q=1$. See inline text for details.}
\label{fit-fig}
\end{figure}

With the present  formalism the Stefan-Bolzmann constant becomes a
function of  i) $q$ (namely, ``$\sigma_q$") and ii) the relevant
range of temperatures. Indeed, for temperatures below $10^{-2}K$
results will be markedly different from those obtained if we
consider $T>1 K$ (see Fig. \ref{fit-fig}). It may be appreciated
that, for these two Temperature ranges, the values of $\sigma_q$
increase monotonically with $q$.

\begin{figure}[tbp]
\begin{center}
\includegraphics[width=11cm]{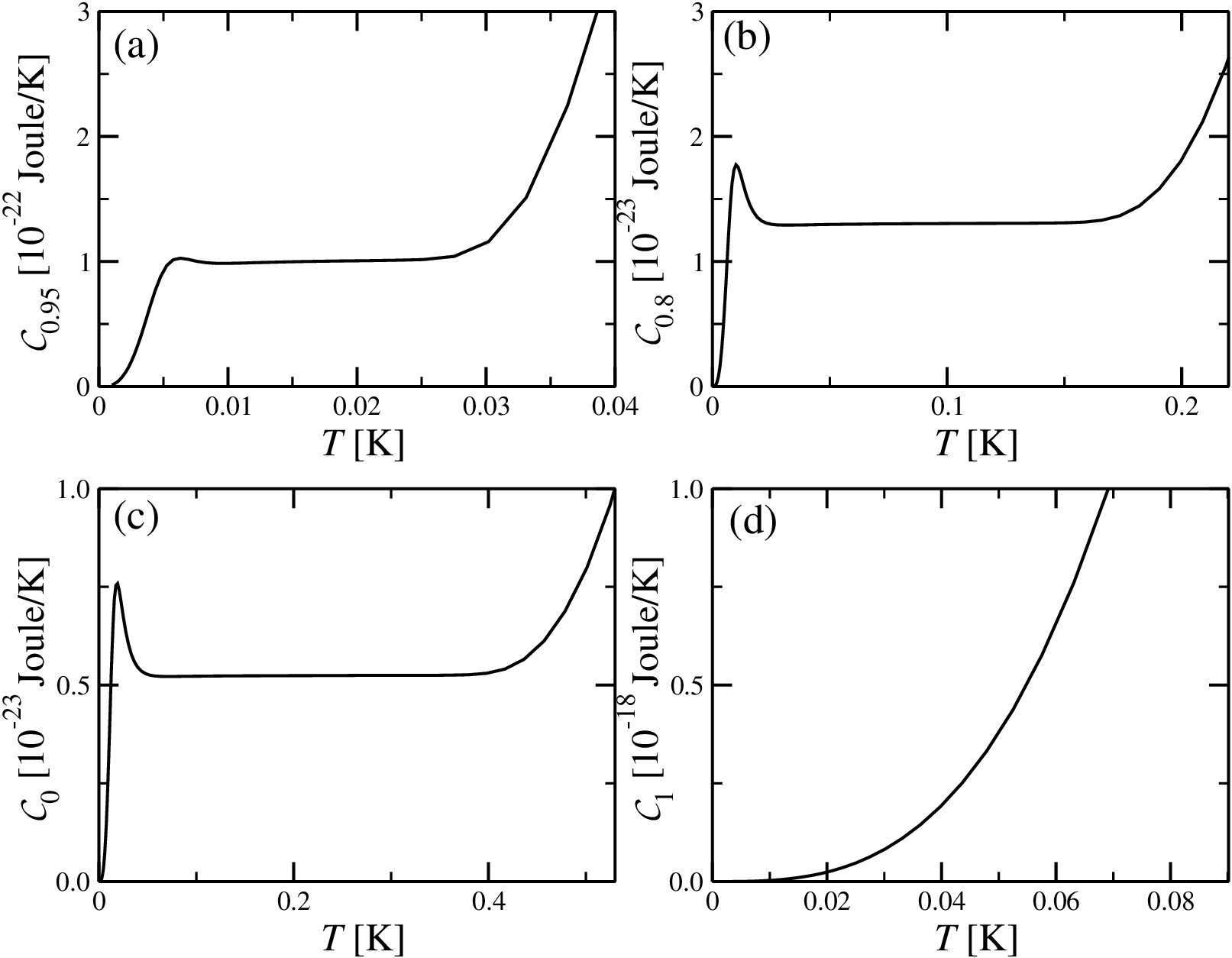}
\end{center}
\caption{Specific heat, ${\cal C}_q$, of a three-dimensional box
of volume $1 \hbox{m}^3$, as a function of Temperature $T$, for
different values of the non-extensivity parameter $q$. The plots
show the arising results for (a) $q=0.98$, (b) $q=0.8$, (c) $q=0$,
(b) $q=1$.} \label{Cq-T}
\end{figure}

Looking for insights into the meaning of the ``violation range"
$10^{-2}K<T<1K$ we considered the specific heat \be
C_q=\frac{dU_q}{dT}. \ee The concomitant results are plotted in
Fig. \ref{Cq-T}. One  sees that the violation range coincides with
that of a constant specific heat, i.e., $U_q\propto T$. The
specific heat curves'  behavior is quite different from that
associated with Gibbs' predictions, even for $q$ values close to
unity. By recourse to detailed numerical analysis  one notes that
the peak one observes is not a discontinuity but part of  a smooth
curve that reflects the typical behavior of a first excited energy
level when that level lies too close to the ground state. The
typical step-like form tends to disappear in the $q\rightarrow 1$
limit, just as  if $1-q$ were a new ``degree of freedom" of the
system.

For more details on the transition between Gibbs' and Tsallis's
statistics let us  analyze the $q\rightarrow 1$ limit in Eq.
(\ref{Uqf}). Using the fact that the pertinent series are highly
convergent, we can try to recover analytically the above numerical
results by keeping only the first order term of the pertinent
series,

\be U_{q} \approx \frac{n\chi_q\xi _{n}}{\beta }\left[
1+(1-q)\beta U_{q}\right] ^{n+1}, \label{Uqfsimp} \ee where
$\chi_q=\Gamma [1/(1-q)] (1-q)^{-(n+1)}\Gamma^{-1} [1/(1-q)+n+1]$.
For $q\rightarrow 1$, $\chi_q\rightarrow 1$ and we see that
Stefan-Boltzmann's law is recovered.

Now, to first order in $1-q$, $$\left[ 1+(1-q)\beta U_{q}\right]
^{n+1} \approx1+(n+1)(1-q)\beta U_{q},$$ so that, rearranging
terms, Eq. (\ref{Uqfsimp}) can be cast as

\be
U_q\approx\frac{n\chi_q\xi _{n}\beta^{-1}}{1-(1-q)(n+1) n\chi_q\xi
_{n}}. \label{Uql1} \ee

The change of behavior we are interested in  is better observed
with reference to Eq. (\ref{Uq2}), by keeping only first order
terms in the series expansion. We find

\be
U_q\approx \frac{n}{\beta}\frac{B_1 \Gamma^{-1}[(2-q)/(1-q)
+n]}{\Gamma^{-1}[(2-q)/(1-q)]+B_1 \Gamma^{-1}[(2-q)/(1-q)+n]}, \label{Uqa1} \ee
where
\be
B_1=\xi_n\frac{[1+(1-q)\beta U_q]^n}{(1-q)^n}. \ee

It is clear that, for $\beta U_q$ such that the first term in the
denominator of Eq. (\ref{Uqa1}) dominates, $U_q\propto T^{n+1}$.
When the second term is dominant, instead, then $U_q\propto T$, in
agreement with our numerical results. Eq. (\ref{Uql1}) displays a
similar behavior, although its validity is restricted to the
$q\rightarrow 1$ limit. This prevents the second term in the
denominator from being dominant, a fact reflected  in Fig.
\ref{Uq-T}, where the linear dependence is seen to fade away.

\subsection{Comparison with previous non-extensive results} \label{history}

The Stefan-Boltzmann's law
was first discussed within the Tsallis' non-extensive framework in
 \cite{cn1}. In this paper the authors  employ the so-called
Curado-Tsallis un-normalized expectation values \cite{t01} and
work in the limit $q\rightarrow 1$: a first order approximation in
$1-q$ is used for the partition function in order to study the
cosmic microwaves' background (in Sect. \ref{cobe} we will apply
the results of the present work to the same data set). In a
subsequent effort, Lenzi {\em et al.} \cite{Lenzi} advanced an
exact treatment for the same problem, also within the
Curado-Tsallis framework. A third relevant work is that of
Tirnakli {\em et al.} \cite{Diego}, that compared the exact
nonextensive treatment of the problem with the FA one, including
  the un-normalized asymptotic approach (AA) already introduced in
\cite{cn1}.

The different ensuing  expressions for the pertinent \`a la
Stefan-Bolzmann laws are given in Table \ref{table} for $n=3$. The
traditional Gibbs expression can be found in the first row, while
the second and third are, respectively, the Curado-Tsallis' and
OLM (exact) results. The TMP solution can be read off the fourth
row, where we have simply used Eq. (\ref{ljTMP}) for $\beta$. The
last three rows are devoted to the $q\rightarrow 1$ limit. The
first one contains the factorization approach result, the second
is the our new, OLM factorization approach expression, and,
finally, the last one gives the  exact OLM treatment in this
limit.

Inspection of these expressions allows one to appreciate that the
OLM-FA seems to inherit characteristics of both the FA and the OLM
treatments for $q\approx 1$. The OLM-FA numerator coincides with
that of the FA approach, while the denominator resembles the OLM
one (except for the sign). Indeed, the concomitant integrals  are
identical here to those of the un-normalized treatment.
\section{Planck and Wien laws}
\label{PWLs}
\subsection{Planck's law: exact OLM treatment}
\label{PL} The generalized spectral energy distribution of
black-body radiation $u_{q}$ is defined by the integral
\begin{equation}
U_{q}=\int\limits_{0}^{\infty }d\omega u_{q}.  \label{uq}
\end{equation}

In order to obtain $u_q$ we need to analyze, again, the trace's
argument in the expression for
 $U_q$. We will follow the path already pursued above, but without
integrating the $q=1$-like term over frequencies. We obtain

\begin{equation}
Tr \left( Z_{1}(\tilde{\beta}) \widehat{H}\right)= \hbar
A_{n}e^{\xi_n}\int\limits_{0}^{\infty }d\omega \omega
^{n}\frac{e^{-\tilde{%
\beta}\hbar \omega }}{1-e^{-\tilde{\beta}\hbar \omega }},
\end{equation}
that, by recourse to the identity
\begin{equation}
\frac{1}{1-e^{-\tilde{\beta}\hbar \omega
}}=\sum\limits_{s=0}^{\infty }\left( e^{-\tilde{\beta}\hbar \omega
}\right) ^{s},
\end{equation}
allows us to cast $U_q$ in the fashion

\begin{equation}
U_q={\mathcal R}_q^{-1}\frac{\Gamma [1/(1-q)]}{2\pi }\hbar
A_{n}\int\limits_{0}^{\infty }d\omega \omega
^{n}\sum\limits_{s=0}^{\infty
}\sum\limits_{m=0}^{\infty }%
\frac{\xi _{n}^{m}}{m!}\int\limits_{-\infty }^{\infty }dt\frac{%
e^{(1+it)[1-(1-q)\beta [\hbar \omega
(1+s)-U_{q}]]}}{(1+it)^{\frac{1}{1-q}%
+nm}}.
\end{equation}

With Eq. (\ref{pi}) the integral above is easily evaluated, and we
obtain
\begin{equation}
U_q={\mathcal R}_q^{-1}\Gamma [1/(1-q)]\hbar
A_{n}\int\limits_{0}^{\infty }d\omega \omega
^{n}\sum\limits_{s=0}^{\infty }\sum\limits_{m=0}^{\infty
}\frac{\xi
_{n}^{m}%
}{m!}\frac{\left[ 1-(1-q)\beta [\hbar \omega (1+s)-U_{q}]\right]
^{\frac{q}{%
1-q}+nm}}{(1-q)^{mn}\Gamma [1/(1-q)+nm]}.
\end{equation}

 According to Eq. (\ref{Cq}), the energy density defined by Eq. (\ref{uq})
is
\begin{equation}
u_{q}(\omega)=\hbar A_{n}\omega
^{n}\frac{\sum\limits_{m=0}^{\infty }B_m S_m \Gamma^{-1}
[1/(1-q)+nm]}{\sum\limits_{m=0}^{\infty }B_m \Gamma^{-1}
[1/(1-q)+nm]},  \label{uqf}
\end{equation}
where $B_m$ is given by Eq. (\ref{Bm}) while  $S_m$ reads \be
S_m=\sum\limits_{s=1}^{s_q }\left[ 1- (1-q)\beta_q \hbar \omega
s\right] ^{\frac{q}{1-q}+nm},  \label{Sm} \ee where \be
s_q=\left[\frac{1}{(1-q)\beta_q \hbar \omega}\right], \ee and \be
\beta_q=\frac{\beta}{1+(1-q)\beta U_q}. \label{betaq} \ee

Notice that a new cut-off condition has been introduced, namely,
$1- (1-q)\beta_q \hbar \omega s>0$, that transforms the original
series into a finite sum.

\begin{figure}[tbp]
\begin{center}
\includegraphics[width=11cm]{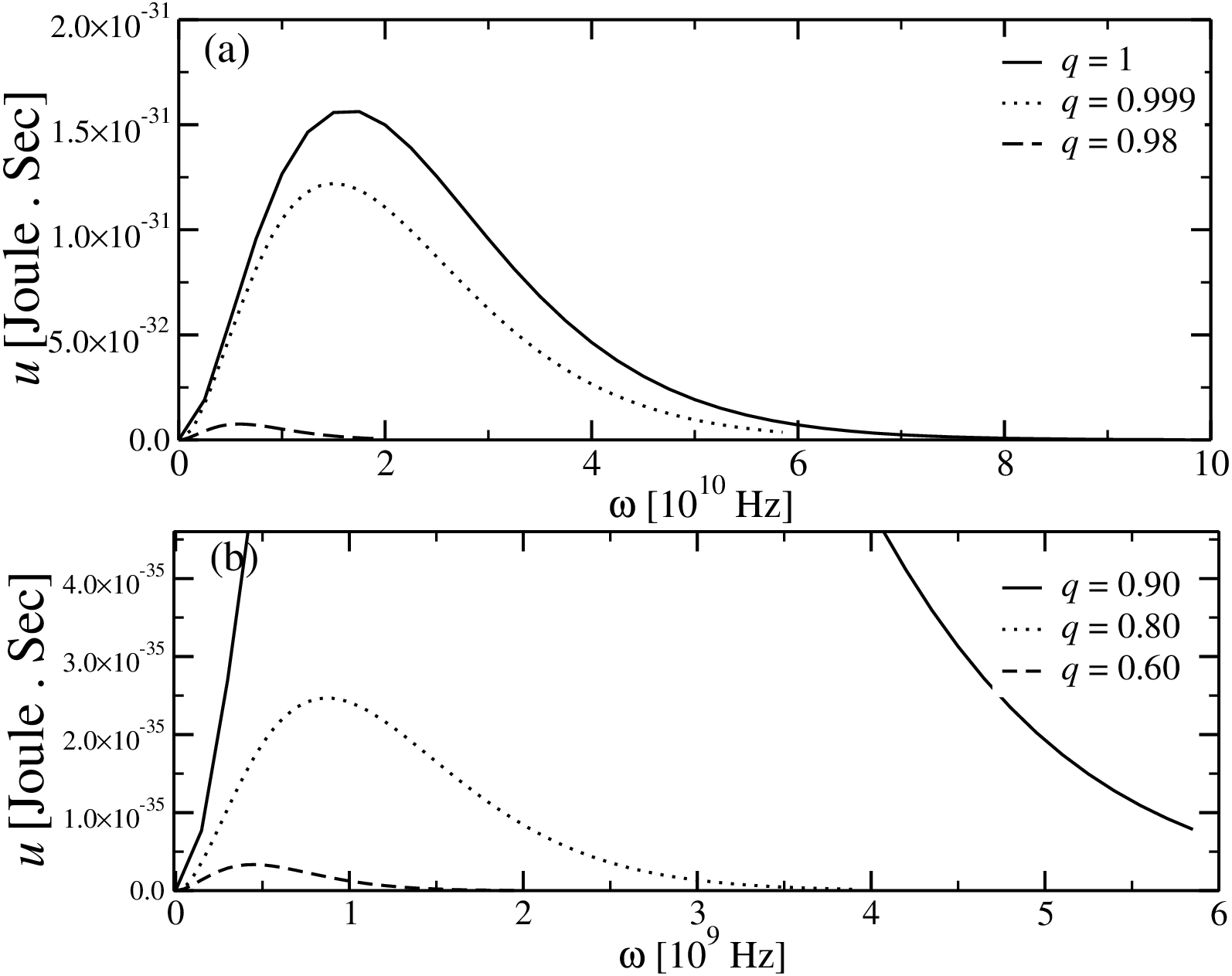}
\end{center}
\caption{Energy density $u_q$ as a function of the frequency
$\omega$. All the curves were obtained for a temperature $T=0.1
K$, and $V=1 \hbox{m}^3$. Inset (a) shows the results for the
limit $q \to 1$, whereas in the inset (b) the results for obtained
otherwise are depicted.} \label{dens-w}
\end{figure}

 Since one knows $U_q$ from Eq. (\ref{Uqf}), an expression for $u_q$ is easily
 obtained. Notice that it is not self referential.
  The shape of the resulting curves is similar to those arising
from the traditional treatment, even for $T$ values where one
detects $q$-violations to  Plank's law.
 Fig. \ref{dens-w} depicts $u_q\;vs\;\omega$ for
different $q$ values. The maximum's values do not coincide with
the ones of the canonical Gibb's  approach when $q$ differs  from
unity. In Fig. \ref{dens-w} (a) it becomes apparent that results
for $q=1$ and for $q=0.98$ greatly differ. The difference seems to
become less important between results for $q=0.9$ and those for
$q=0$, as can be seen in (b). Another point to be stressed is that
the maximum's position will also depend on $q$, not just on $T$,
as in the traditional Wien's law.

We have seen from Fig. \ref{dens-w} that Eq. (\ref{uqf}) yields
results that quite resemble the ones given by Planck's law. It is
then reasonable to look for a perturbative expansion in $1-q$. The
series over the $m$ factor in Eq. (\ref{uqf})  rapidly converges.
Only the first terms are important. In the limit $q\rightarrow 1$,
the exponent $q/(1-q)+nm\approx q/(1-q)$ in Eq. (\ref{Sm}), and
the finite sum becomes a series expansion. We have

\be u_q\approx\hbar A_{n}\omega ^{n}S, \label{uqS} \ee where \be
S=\left.\sum\limits_{s=1}^{\infty } \left[ 1- (1-q)\beta_q \hbar
\omega s\right] ^{\frac{q}{1-q}}\right|_{q\rightarrow
1}\approx\sum_{s=0}^{\infty}\exp\left[ - q \beta_q \hbar \omega
(1+s) \right], \ee a power series in $s$ of guaranteed
convergence, i.e.,

\begin{equation}
S\approx \frac{1}{e^{ q\beta_q \hbar \omega}-1},  \label{Smap}
\end{equation}
so that replacement into Eq. (\ref{uqS}) yields

\begin{equation}
u_q\approx \frac{\hbar A_n \omega^n}{e^{ q\beta_q \hbar
\omega}-1}, \label{uqapp}
\end{equation}
a first order correction to the Planck law. The classical result
is attained for $q\rightarrow 1$. Eq. (\ref{uqapp}) provides one,
then, with an approximate energy density.

Following the standard text-book treatment  \cite{Huang}, we can
define the particle-density as

\be n_q= (\hbar A_n \omega^n)^{-1} u_q, \ee which, from Eq.
(\ref{uqapp})   can be written, to first order in  $1-q$ as

\be n_q= \frac{1}{e^{ q\beta_q \epsilon}-1}, \label{nq} \ee with
$\epsilon=\hbar \omega$.

We have encountered  an alternative expression for the boson
particle-density of the black-body radiation, to be compared to
the ones that result from either the  FA or OLM-FA (see Apendix
\ref{OLMFA}). The correct  value for the $q \rightarrow 1$ limit
is obtained.

  Consider now the specific heat curves of Sect.
\ref{comp}. From Eq. (\ref{betaq}) we see that, to first order in
$1-q$ we have

\be \beta_q\approx \beta[1-(1-q)\beta U_q], \ee that, replaced
into Eq. (\ref{nq}) and using the fact that $q=1-(1-q)$ leads to
(keeping only terms of order $1-q$)

\be n_q\approx\frac{1}{e^{ \beta (\epsilon-\mu_q^*)}-1},
\label{nq1} \ee where \be \mu_q^*=(1-q)\epsilon (1+\beta U_q) \ee
plays the role of a fictitious $q$-chemical potential, such that
$\mu_q^*\rightarrow 0$ for $q\rightarrow 1$. Of course, the
``true" {\em physical} chemical potential vanishes identically
(see Sect. \ref{cn}). A sort of fictitious $q$-Bose-Einstein
condensation effect might seem to be implied by the presence of
this pseudo-chemical potential, a point that  deserves further
careful exploration, to be addressed in a future work.

\subsection{Wien law}
\label{WiL} A nice result of the traditional thermostatistics is
the  linear relation  that exists between the frequency $\omega^*$
that maximizes the energy density spectrum, on the one hand, and
the temperature on the other one
\begin{equation}
\omega ^{*}\propto T.
\end{equation}
This relation is known as Wien's law.

 According to  the results depicted in
  Fig. \ref{dens-w} (see Sect. \ref{PL}),
 deviations of the these maximum values from what we expect from
them
 according to the above relation will depend on $q$.
If  we perform the pertinent derivatives in Eq. (\ref{uqf}) we
obtain

\be \label{WL}
\left.\sum\limits_{m=0}^{\infty}B_m\Gamma^{-1}\left[1/(1-q)+nm\right]
\left( \frac{n}{\omega}S_m+\frac{dS_m}{d\omega}
\right)\right|_{\omega^*}=0, \ee where $S_m$ depends in a non
trivial manner on $\omega$ due to the cut-off condition (see Eq.
(\ref{Sm}): the cut-off condition ``cuts" the series at an  $s_q$
that depends on $\omega$).

\begin{figure}[tbp]
\begin{center}
\includegraphics[width=11cm]{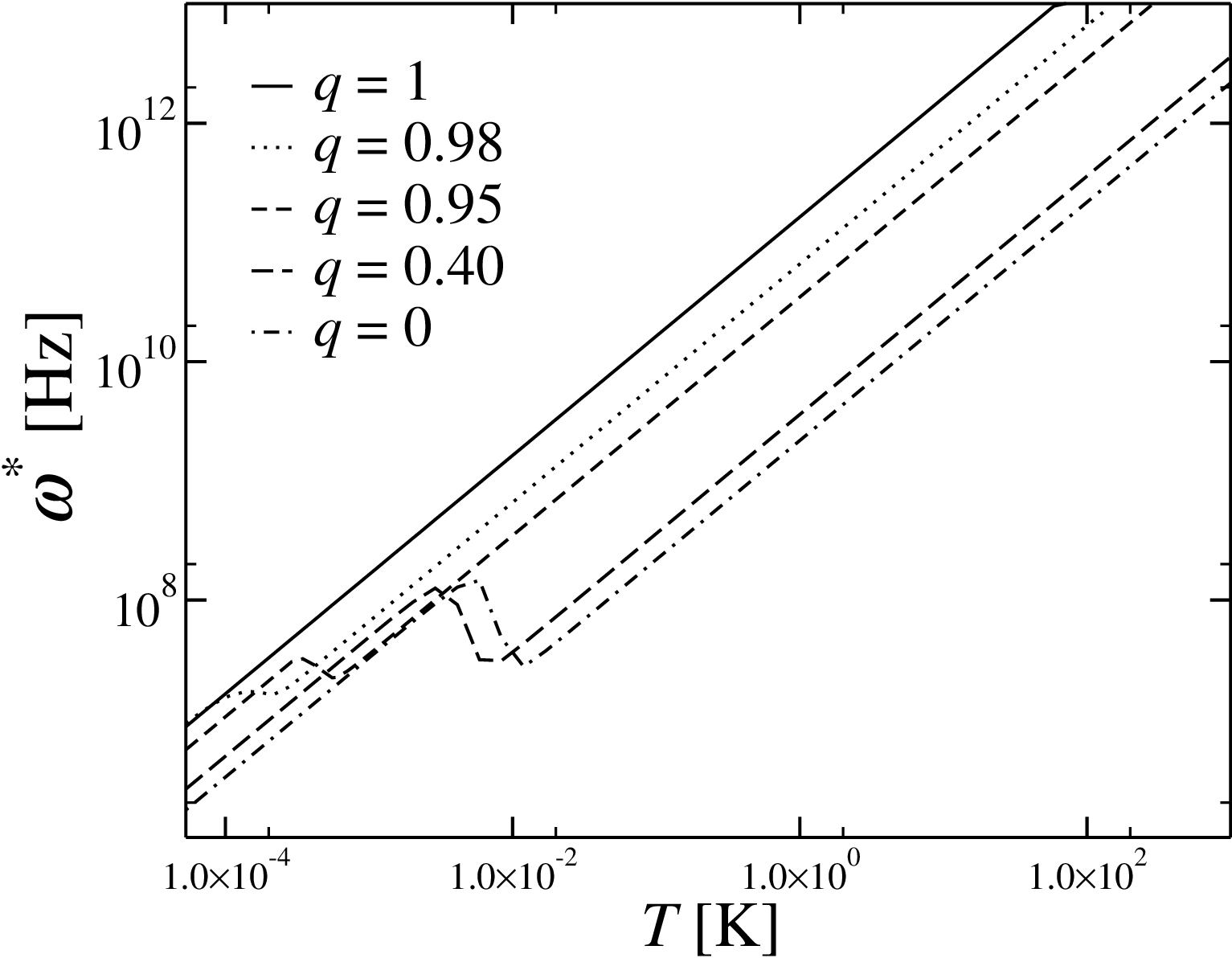}
\end{center}
\caption{Frequency $\omega^*$ for which the energy density reaches
its maximum, as a function of temperature $T$, for different
values of $q$. As in the previous figures, $n=3$. } \label{fWien}
\end{figure}

Numerical calculations of Eq. (\ref{WL}) are depicted in Fig.
\ref{fWien}. It can be seen that the linear dependence is
respected over temperatures above $10^{-2} K$ and below $10^{-4}
K$. There is a ``violation range"  the $T$-zone
($10^{-4}<T<10^{-2}$). This implies  lower values of $T$ than
those that violate Stefan Boltzmann's  law. Departures from Wien's
law are observed even for $q-$values as close to unity as
$q=0.98$. Using now Eq. (\ref{uqapp}) as a starting point we also
get first order corrections in $1-q$ to the Wien law. We can
locate the maximum with
respect to $%
\omega$ of Eq. (\ref{uqapp}) with the auxiliary definition
\begin{equation}
x=\frac{\omega q \beta \hbar}{1-(1-q) \beta U_q},  \label{x}
\end{equation}
so that we immediately find

\begin{equation}
e^{-x}+\frac{x}{n}=1,
\end{equation}
whose solution is a constant,  $b$, for each fixed value of $n$.
For $n=3$ we find, for instance, $b=2.82$. The maximum of
$u_{q}(\omega )$ for different $T$'s is located at distinct
frequencies according to

\begin{equation}
\omega _{i}=\frac{bk}{q\hbar
}T_{i}-\frac{1-q}{q}\frac{U_{q}}{\hbar },
\end{equation}
a first order correction to Wien's law.


\subsection{Microwave Cosmic Background Radiation}
\label{cobe}

The OLM formalism can be applied to the analysis of the cosmic
microwave radiation in order to search for putative deviations
from Planck's law. The best source of data on this topic comes
from  the COBE satellite (Far-infrared Absolute Spectrophotometer
(FIRAS)). The ensuing brightness, when compared to  a Planck
spectrum at the temperature  $T=2.72584 \pm 0.0005 K$ shows
significant deviations from the expected form (the $\chi^2$ of the
fit to a Planck spectrum is rather large).

FIRAS measures the {\it differential} spectrum between the cosmic
background and an internal reference adjusted to  approximately
$2.7 K$ so as to i) avoid problems with imperfect emissivity and
ii)  limit the dynamic range of the instrument. In addition, an
external black-body is used to calibrate (according to Gibbs'
statistics) the gain of the instrument with a temperature in the
range of $ 2K$ up to $25K$.

\begin{figure}[tbp]
\begin{center}
\includegraphics[width=11cm]{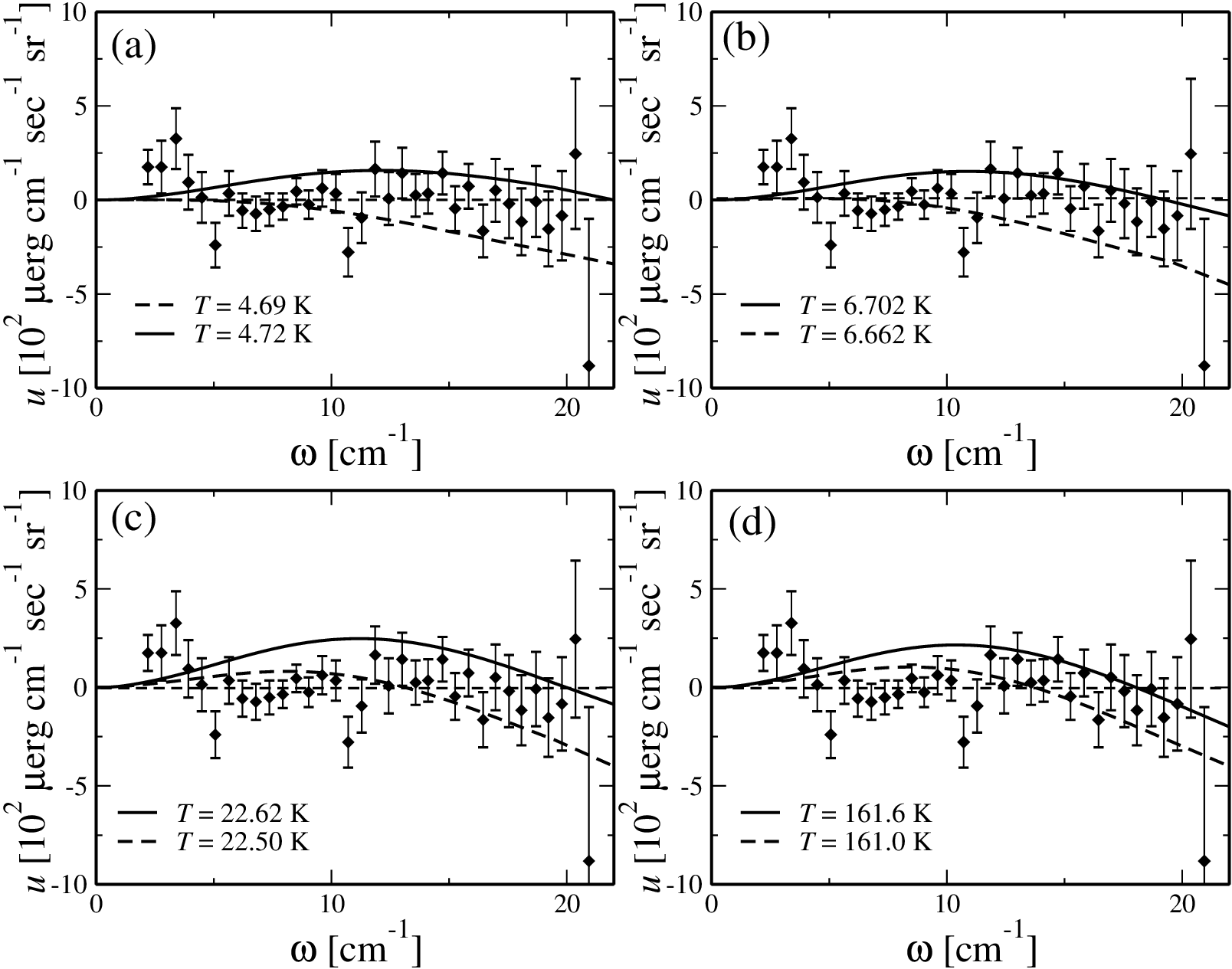}
\end{center}
\caption{We show the differential radiation between for the FIRAS
spectrographer onboard COBE satellite. Superimpossed with the
experimental data the best fits for four different $q$ values are
also shown.} \label{cobe-fit}
\end{figure}

This is the kind of data that will be of interest in order to
verify the predictions of the OLM formalism. If we assume the
cosmic background does not obey the usual Planck's law, but the
OLM one, we can check the concomitant differences and contrast the
ensuing values with those obtained by FIRAS. The results are
depicted in Fig. \ref{cobe-fit}. We plot the best fits for four
different values of $q$. The range of temperatures that allow for
 the best fitting is also given. In all the graphs it is apparent that
deviations are well
  predicted if we assume a statistics with $q \neq 1$ for the cosmic
  background. As $q$ deviates from unity higher temperatures are
predicted, in agreement  with the results displayed in Fig.
\ref{dens-w}. We see there  that the spectrum maximum's value
becomes smaller for a given, fixed temperature when $q$ deviates
from unity.

The experimental data depicted in Fig. \ref{cobe-fit} seems to
imply that the statistics ruling the cosmic microwave radiation is
different from that of the conventional black-body instance. This
constitutes evidence for the existence of alternative statistics,
and tells us that the statistics behind these phenomena seems to
possess  an additional, $q$-degree of freedom.

\section{Conclusions}

\label{conclusiones} Within the $q$-Thermostatistics framework we
perform an exact analysis of the $n$-dimensional black-body
radiation process. We employ to such effect both  Tsallis' and
R\'enyi's entropies, within the range $0<q<1$. The new theoretical
ingredient here is the so-called OLM approach to non-extensive
thermostatistics \cite{OLM}.

 We develop a $q$ generalization of several laws:
Stefan Boltzmann's , Planck's, and Wien's.  We find the the
conventional behavior still obtains for temperatures above $1 K$,
although there is  a $q$ dependence in the appropriate
proportionally constants. We recover the traditional relationship
between radiation pressure and internal energy using the OLM
formalism.

 We apply the formalism
 to experimental data on the cosmic microwaves'
background and  reproduce it with acceptable accuracy even for
$q$-values that appreciably differ from unity. The larger $|1-q|$
is, the  higher the predicted equilibrium cosmic temperature.

\smallskip \acknowledgements The financial support of the National
Research Council (CONICET) of Argentina is gratefully
acknowledged. F. Pennini acknowledges financial support from UNLP,
Argentina. S. Mart\'\i nez wants to thank R. Rosignoli for fruitful
discussion.

\appendix

\section{The OLM Factorization Approximation}
\label{OLMFA}
\subsection{Un-normalized Factorization Approximation (FA)}

In order to facilitate the reader's task we review now the
factorization approach (FA) treatment, due to  B\"uy\"ukkilic {\em
et al.}, who tackled in Ref. \cite{FA} the quantum ideal gas
problem in a grand canonical scenario using the CT-Tsallis
formalism. Tsallis' entropy is thereby maximized subject to the
un-normalized constraints

\begin{eqnarray}
Tr \rho^q \hat H & =& U_q\\ Tr \rho^q \hat N &=&  N_q.
\label{vincs1}
\end{eqnarray}
 The pertinent partition function is

\be Z_q^{CT}=\sum_i [1-(1-q) \sum_j n_{ij} x'_j
]^{1/(1-q)}\label{Zq1}, \ee where $n_{ij}$ are the occupation
numbers of the level $j$ (with single particle energy
$\epsilon_j$) for a given $i$-microscopic  configuration and \be
  x'_j = \beta^{CT} (\epsilon_j -\mu^{CT}), \label{xj} \ee
with $\beta^{CT}$ the inverse temperature and $\mu^{CT}$ the
chemical potential.

According to \cite{FA}, ``there is no restriction on the summation and 
thus one an factorise a product of factors, one for each one-particle state,
since particle are regarded statistically independent"
. Therefore
(the essential point of the FA approach), the partition function
$Z_q^{CT}$ given by Eq. (\ref{Zq}) can be factorized. Each factor
 corresponds to a single particle state

\be Z_q^{CT}\approx\prod_{j=0}^{\infty} \sum_{i=0}^w
\left[1-(1-q)n_{ij}x'_j \right]^{1/(1-q)}. \label{ZqCTFA} \ee

The average occupation number $\left<n_j\right>_{FA}$ of the state
$j$ becomes then \cite{FA} \be \left<n_j\right>_{FA}\approx
\frac{1}{[1-(1-q)x'_j]^{1/(1-q)} \mp 1}, \ee which is a
non-extensive, FA-Tsallis generalization of the Bose-Einstein
(Fermi-Dirac) distribution.

\subsection{The normalized OLM treatment}
\label{NFA} An alternative treatment to the factorization approach
using the TMP formalism yields self-referential probabilities. The
OLM procedure, instead, is not afflicted by such a problem. One
again maximizes Tsallis' generalized entropy given by Eq.
(\ref{entropia}) subject to the diagonal normalized constraints
for the Grand Canonical Ensemble, which results in  a partition
function of the form

\begin{equation}
\bar{Z}_{q}=e_q^{x_q}\sum_{i}\left[1-(1-q) \sum_j n_{ij}
x_j\right]^{1/(1-q)},  \label{Zq2}
\end{equation}
where $e_q^x=[1+(1-q) x]^{1/(1-q)}$ and, instead of $x'_j$ (Eq.
(\ref{xj})), we have

\be \label{newx} x_j =
  \frac{\beta(\epsilon_j-\mu)}{1+(1-q) x_q} ,
\ee with \be x_q = \beta (U_q - \mu N_q).\label{x'j} \ee

Note that $x_j \rightarrow x_j^{\prime}$ for $q \rightarrow 1$.
The conventional limit is the same for both treatments, if one
remembers that the un-normalized methodology uses a partition
function $Z_q^{CT}$, while the normalized treatment deals with a
{\em bar} partition function $\bar{Z}_q$ (see Eq. (\ref{lnqz'})).
The concomitant $q=1$-limits are related according to
 $Z_1=e^{-\beta(U-\mu N)}\bar{Z}_1$.

Proceeding once again as in Ref. \cite{FA}, i.e., neglecting
correlations between particles and regarding states of different
particles as statistically independent, the partition function
$\bar{Z}_q$ can be factorized. Each factor corresponds to a single
particle state,

\be \bar{Z}_q\approx e_q^{x_q}\prod_j \sum_i\left[1-(1-q)n_{ij}x_j
\right]^{1/(1-q)}. \ee

The present expression for $\bar{Z}_q$ is {\it formally} identical
to that encountered in Ref. \cite{FA} (see Eq. (\ref{ZqCTFA})),
save for a multiplicative factor, which does not contribute to the
ensuing microscopic probabilities. Thus, according to Ref.
\cite{FA}, the average value $\langle n_j\rangle_{NFA}$ is now

\be \langle
n_j\rangle_{NFA}\approx\frac{1}{[1-(1-q)x_j]^{1/(1-q)} \mp 1}
\label{nNFA} \ee with $x_j$ given by Eq. (\ref{x'j}).

We introduce here the normalized factorization approach (NFA)
black-body's treatment. The mean occupation number is
 given by Eq. (\ref{nNFA}) with $\mu=0$ (the chemical potential vanishes because we are working with photons).
In the continuum limit $\epsilon(\omega)=\hbar \omega$ ($\omega$
is the frequency), and the particle-density will be given by

\be \left<n(\omega)\right>_{NFA}\approx
\frac{1}{[1-(1-q)x'(\omega)]^{1/(1-q)}- 1}, \ee with \be
 x'(\omega)=\frac{\beta \hbar \omega}{1+(1-q)\beta U_q}.
\ee

In the three dimensional case the mean occupation number is
connected with the energy-density according to

\be u_q(\omega)=\frac{\hbar V}{\pi^2 c^3}
\omega^3\left<n(\omega)\right>_q. \label{NFAuq} \ee

Keeping only first order terms in $1-q$ in Eq. (\ref{NFAuq}),
integrating over frequencies, and re-arranging terms, the
internal energy reads

\be U_q^{NFA}=\sigma T^4 \frac{1-
(1-q)\theta}{1+\frac{4}k(1-q)\sigma T^3}, \ee where
$\sigma=(\pi^2 k^4 V)/(15 \hbar^3 c^3)$ (with $c$, the light's
speed) is in closed connection with the Stefan-Boltzmann constant
$(c\sigma)/(4V)$ \cite{Reif}.

%
%

\begin{table}
\begin{tabular}{||l|c|c||}
\hline Formalism & Result & \\ \hline \hline
 Stefan-Boltzmann' law & \( U_1=\sigma T^4 \) & $\sigma=(\pi^2 k^4
V)/(15 \hbar^3 c^3)$ \\
 \hline
C-T Solution & \(
U_q^{CT}=3\xi_3kT\Gamma^{1-q}\left[\frac{2-q}{1-q}\right]
\frac{\sum\limits_{m=0}^{\infty} \frac{\xi_3^m}{m!}
\Gamma^{-1}\left[\frac{2-q}{1-q}+ 3m+3\right] }{\left[
\sum\limits_{m=0}^{\infty} \frac{\xi_3^m}{m!}
\Gamma^{-1}\left[\frac{2-q}{1-q}+ 3m\right] \right]^q} \) & \(
\xi_3=\frac{4\Gamma(3)\zeta(4)V}{\Gamma(3/2)}
\left(\frac{\sqrt{\pi} k T}{h c(1-q)}
 \right)^3\)\\
 \hline
OLM Solution& \(U_{q}^{OLM,R}=3\xi _{3}kT\frac{\left[ 1+(1-q)\beta
U_{q}\right] ^4}{(1-q)^4}\frac{\sum\limits_{m=0}^{\infty} B_m
\Gamma^{-1}
\left[\frac{1}{1-q}+3m+4\right]}{\sum\limits_{m=0}^{\infty }B_m
\Gamma^{-1} \left[\frac{1}{1-q}+3m\right]}\)& \(B_m=\frac{\xi
_{3}^{m}}{m!}\frac{\left[ 1+(1-q)\beta U_{q}\right]
^{nm}}{%
(1-q)^{3n}}\)\\
 \hline
 \hline
FA Aproximation & \(U_q^{(FA)}\approx\sigma [1-(1-q)\theta] T^4\)
& \( \theta=\zeta(5) \Gamma(6)/(2 \zeta(4) \Gamma(4)) \) \\

OLM-FA Aproximation& \(U_q^{NFA}\approx\sigma \frac{1-
(1-q)\theta}{1+4 (1-q)\sigma T^3/k} T^4 \) & \\ \hline OLM
Aproximation& \( U_q^{OLM}\approx \sigma\frac{1}{1-4(1-q)\sigma
T^3/k }T^4
 \) & \\
\hline \hline
\end{tabular}
\caption{ The results of different generalizations of
Stefan-Boltzmann's law.}
\end{table}
\label{table}
\end{document}